\documentclass[12pt]{iopart}
\usepackage{iopams}  
\usepackage{psfrag,graphicx,color,float,cite}
\def\bea{\begin{eqnarray}}
\def\eea{\end{eqnarray}}

\newcommand{\qed}{\nobreak \ifvmode \relax \else
      \ifdim\lastskip<1.5em \hskip-\lastskip
      \hskip1.5em plus0em minus0.5em \fi \nobreak
      \vrule height0.75em width0.5em depth0.25em\fi}

\begin{document}

\title[On the construction of partial difference  schemes II]{On the construction of partial difference  schemes II: discrete variables and Schwarzian lattices. \\}
\author{Decio Levi$^1$ and Miguel A. Rodr\'{\i}guez$^2$}
\address{$^1$Dipartimento di Matematica e Fisica, Universit\`a degli Studi Roma Tre and INFN Sezione di Roma Tre, Via della Vasca Navale 84, 00146 Roma, Italy  \\ $^2$Departamento de F\'{\i}sica Te\'{o}rica II, Facultad de F\'{\i}sicas, Universidad Complutense, 28040 Madrid, Spain}
\ead{decio.levi@roma3.infn.it, rodrigue@fis.ucm.es}



%

\begin{abstract}
In the process of   constructing  invariant  difference  schemes  which
approximate  partial  differential  equations we write down a procedure for discretizing an arbitrary  partial differential equation on an arbitrary lattice.  An open  problem  is the meaning of a lattice which does not satisfy   the  Clairaut--Schwarz--Young  theorem.   To analyze it we  apply  the  procedure  on  a
simple example, the potential Burgers equation with two different lattices, an
orthogonal lattice which is invariant under the symmetries of the equation and satisfies
the commutativity of the partial difference operators and an exponential lattice which
is not invariant and does not satisfy the  Clairaut--Schwarz--Young  theorem.  A discussion on the numerical
results is also presented showing the different behavior of both schemes for two different
exact solutions and their numerical approximations.
\end{abstract}



\submitto{Acta Polytecnica}
\vspace*{0.1cm}\begin{center}
{\it Dedicated to our Masters Pavel Winternitz and Ji\v{r}\'{\i} Patera on the occasion of their 80 years birthday.}
\end{center}

\section{Introduction}

The construction of difference  equations, written as  invariants of the continuous group of symmetries of differential equations, is part of a project to apply symmetry group methods to the numerical solution of differential equations  \cite {d91,doro2,DW00,win2,win1,lw2006,LOTW,leviscimiternathomovawinternitz2012,levithomovawinternitz2011,LW1996,LW91,rv13,RebeloValiquette,rw09,W2004,PW8,ko2004}. This project has accomplished a significative advance since its first introduction  last century.  In particular, the construction of invariant schemes has proven to be a very fruitful approach in the construction of numerical schemes for ordinary differential equations \cite{win1,win2}, in cases when the usual approaches present serious problems of convergence and accuracy,  for instance, in the behavior of the solutions in the neighborhood of a singularity. In this problem a deeper understanding of the mathematics involved in its relation with numerics can provide results important for their applications to problems in Physics and Mathematics.

The usual procedure in this framework is to compute the symmetry group of the differential equation and then compute the invariant lattice and invariant difference equation with respect to the symmetry  group. However, since the  differential and difference calculus present substantial differences, a special care has to be taken to assure the consistency of the approach. For example, the Clairaut--Schwarz--Young theorem on the equality of the cross derivatives, which is satisfied in the continuous case under some mild conditions on the functions, is not valid in the discrete case for a general lattice. Recently it has been shown \cite{LRI} that  the discrete Clairaut--Schwarz--Young theorem, equality of the cross differences, imposes strong restrictions on the lattice. Moreover, the construction of the discrete invariant scheme starting from the discrete invariants is not at all obvious as it is usually obtained by finding a proper combination of the continuous limits of the various discrete invariants. The main idea guiding the construction of the whole scheme, that is the difference equations and the equations defining the lattice, which in some cases are mixed, is that the continuous limit yields the differential equation and trivial identities.

This article is a continuation of our work on the construction of partial difference schemes \cite{LRI}.  In the previous work we concentrated on the Clairaut--Schwarz--Young theorem. Here we  introduce by a one to one correspondence a  new set of discrete coordinates  which describe the partial difference equation on the lattice. In terms of these coordinate systems we can write immediately the discrete counterpart of any continuous invariant. So we can discretize in a  straightforward way any partial differential equation
described in terms of the invariants of a group of symmetries.

A particular role in the construction of discrete invariant schemes is played by the lattice. Consequently the Clairaut--Schwarz--Young theorem can play an important role in discriminating compatible lattice schemes, i.e. the combination of the discrete equation and its lattice.

In  Section 2  we study schemes for scalar partial differential equations and show the constraints on the group transformations due to the Clairaut--Schwarz--Young theorem. Using these results, in Section 3 we construct in a standard way the invariant discrete potential Burgers and in Section 4, we study it numerically for two different lattices, one invariant and Schwarzian and one not, for two different exact solutions of the differential equation.  Some concluding remarks are presented in Section 5.


\section{Schemes for partial difference equations}

In the case of ordinary difference equations of order $K$ for one dependent variable $u_n$ and one independent variable $x_n$, a natural scheme is given by  the points $ \{ x_{n+k-1},u_{n+k-1}, 1 \leq k \leq K+1 \} $ for some fixed $n$.
An alternative equivalent set of coordinates on the scheme is given by \cite{PW8,LRI}:
\bea
\{x_n,u_n,p_{n+1}^{(1)},p^{(2)}_{n+2},p^{(3)}_{n+3},p^{(4)}_{n+4}, \ldots p^{(K-1)}_{n+K-1},p^{(K)}_{n+K},h_{n+1},h_{n+2},
\ldots h_{n+K}\}  \label{e3-7}
\eea
with
\bea
h_{n+k}&=&x_{n+k}-x_{n+k-1}, \qquad 1\le k \le K\nonumber \\
p^{(k)}_{n+k}&=&[D_x]^k u_n, \; \mbox{where} \quad D_x = \frac{1}{h_{n+1}}[T_x-1], \; T_x u_n = u_{n+1}, \quad k \in \mathbb Z^+ \label{e3-8} 
\eea
In the continuous limit $h_{n+k} \rightarrow   0$ and $
p^{(k)}_{n+k} \rightarrow \frac{d^k u(x)}{d x^k}$.
If we transform the standard {\it discrete} prolongation \cite{lw2006}
\bea
\mbox{pr}^{(K)} X = \sum_{k=n}^{n+K}  \left(  \xi_k (x_k, u_k)\partial_{x_k} + \phi_k (x_k, u_k) \partial_{u_k} \right) \label{e3-5}
\eea
 of the vector field
 \bea
X = \xi_n(x_n, u_n) \partial_{x_n} + \phi_n (x_n, u_n) \partial_{u_n} \label{e3-4}
\eea
 to the new
variables (\ref{e3-7}), we obtain
\bea
\mbox{pr}X&=&\xi_n (x_n, u_n) \partial_{x_n} + \phi_n (x_n, u_n) \partial_{u_n} +
\sum_{k=1}^{K} \kappa^{(k)} \partial_{h_{n+k}} +
\sum_{k=1}^{K} \phi_{n+k}^{(k)} \partial_{p^{(k)}_{n+k}} \label{e3-10}
\eea
The general formulas for the coefficients $\kappa^{(k)}$ and $\phi^{(k)}$
are
\bea
\kappa^{(k)} &=& \xi_{n+k} - \xi_{n+k-1},      \label{e3-11}   \\
\phi^{(k)}_{n+k}&=&D_x \phi^{(k-1)}_{n+k-1} - p^{(k)}_{n+k}   D_x \xi_{n}  \quad
 k  =  1,2,\cdots , K.  \nonumber
\eea
It is worthwhile to notice that both the higher order discrete derivatives and their corresponding prolongations are written in terms of $x_n$, $u_n$, $\xi_n(x_n, u_n)$, $\phi_n(x_n, u_n)$ and their difference consequences obtained by applying the operator $D_x$ given in (\ref{e3-8}). This way of constructing  invariant ordinary difference equations  is easily extendible to the case of partial difference equations.

For partial difference equations we consider here only, for simplicity,  the case of one dependent variable and two independent variables as will be the example we will discuss in the following Section.  Moreover, as we will deal with a nonlinear partial difference equation of second order we limit ourselves to a scheme of six points ($n,m$), ($n+1,m$), ($n,m+1$), ($n+2,m$), ($n,m+2$), ($n+1,m+1$), the minimum number of points necessary to get all partial second derivatives as first order approximations. The variables $x$, $y$ and $u(x,y)$ in all points correspond to  18 data, 12 related to the independent variables and 6 to the dependent one. The extension to more variables and higher order equations requires just more points but it is straightforward.
Having 12 data for the independent variables we can construct from them 10 differences 
\bea \nonumber
\fl x_{0,0}, \,y_{0,0},\quad &&h^x_{0,0}=x_{1,0}-x_{0,0}, \,h^y_{0,0}=y_{0,1}-y_{0,0}, \, \sigma^x_{0,0}=x_{0,1}-x_{0,0}, \, \sigma^y_{0,0}=y_{1,0}-y_{0,0}, \\ \nonumber
&&h^x_{1,0}=x_{2,0}-x_{1,0},\, h^y_{0,1}=y_{0,2}-y_{0,1},\, \sigma^x_{0,1}=x_{0,2}-x_{0,1}, \, \sigma^y_{1,0}=y_{2,0}-y_{1,0}, \\ \label{f_1}
&& h^x_{0,1}=x_{1,1}-x_{0,1}, \quad h^y_{1,0}=y_{1,1}-y_{1,0},
\eea
where for convenience of notation here and in the following, whenever it may not create misunderstanding, we have indicated just the distance from the values  $n,m$ of the indices.
From the values of the dependent variables in the 6 points we can calculate the 6 quantities
\bea \label{f_2}
u_{0,0}, \; D_x u_{0,0}, \; D_y u_{0,0}, \; [D_{x}]^2 u_{0,0}, \; [D_{y}]^2 u_{0,0}, \; D_x D_{y} u_{0,0}, 
\eea
where the operators $D_x$ and $D_y$,  introduced in \cite{LRI}, are given by
\begin{equation}\label{diff}\eqalign{
{ D}_x=\frac{1}{h^x_{0,0}h^y_{0,0}-\sigma^x_{0,0}\sigma^y_{0,0}} (h^y_{0,0}\Delta_n-\sigma^y_{0,0}\Delta_m) \\
{ D}_y=\frac{1}{h^x_{0,0}h^y_{0,0}-\sigma^x_{0,0}\sigma^y_{0,0}}(-\sigma^x_{0,0}\Delta_n+h^x_{0,0}\Delta_m )}
\end{equation}
with 
\begin{equation*}
T_{n}f_{n,m}=f_{n+1,m},\quad \Delta_n=T_n-1,
\end{equation*}
and 
\begin{equation*}
T_{m}f_{n,m}=f_{n,m+1},\quad \Delta_m=T_m-1.
\end{equation*}

These operators are the discrete counterpart of the partial derivatives in the $x,y$ axes directions, written in terms of the partial difference operators in the lattice directions, which, generically, do not coincide with the cartesian axes. See \cite{LRI} for a more detailed description.

Note that $D_y D_{x} u_{0,0}$ is not independent from the 6 quantities (\ref{f_2}). It  can be written in term of (\ref{f_1}) and (\ref{f_2}). For a generic lattice we have:
\bea \label{f_2a}
\fl D_y D_{x} u_{0,0} =-D_x D_{x} u_{0,0} \frac{\left(h^x_{0,0}-h^x_{0,1}\right) \left(h^x_{0,0}-\sigma^x_{0,0}\right)}{h^x_{0,1} h^y_{0,0}+h^x_{0,0} \sigma^y_{0,0}-h^x_{0,1} \sigma^y_{0,0}-\sigma^x_{0,0} \sigma^y_{0,0}}+ \\ \nonumber \fl \qquad \qquad +D_y D_{y} u_{0,0}\frac{ \left(h^y_{0,0}-h^y_{1,0}\right) \left(h^y_{0,0}-\sigma^y_{0,0}\right)}{h^x_{0,1} h^y_{0,0}+h^x_{0,0} \sigma^y_{0,0}-h^x_{0,1} \sigma^y_{0,0}-\sigma^x_{0,0} \sigma^y_{0,0}}+ \\ \nonumber \fl \qquad \qquad+D_x D_{y} u_{0,0} \frac{h^x_{0,0} h^y_{1,0}+h^y_{0,0} \sigma^x_{0,0}-h^y_{1,0} \sigma^x_{0,0}-\sigma^x_{0,0} \sigma^y_{0,0}}{h^x_{0,1} h^y_{0,0}+h^x_{0,0} \sigma^y_{0,0}-h^x_{0,1} \sigma^y_{0,0}-\sigma^x_{0,0} \sigma^y_{0,0}}
\eea

Formulas (\ref{f_1}--\ref{f_2a}) can be simplified if  we  require the validity of  the Clairaut--Schwarz--Young theorem i.e. $D_x D_{y} u_{0,0}=D_y D_{x} u_{0,0}$.  This is true when the following constraint for the lattice holds \cite{LRI}:
\bea  \label{f_5}
\sigma^x_{n,m} &=\sigma^x_{n+1,m}\equiv \sigma_m^x, \qquad h^x_{n,m}&=h^x_{n,m+1}\equiv h^x_n, \\ \nonumber
\sigma^y_{n,m} &=\sigma^y_{n,m+1}\equiv \sigma^y_n, \qquad h^y_{n,m}&=h^y_{n+1,m}\equiv h^y_m.
\eea

Using the operators $D_x$ and $D_y$ given in (\ref{diff})  
we can transform the  standard {\it discrete} prolongation \cite{lw2006}
\bea
\fl \mbox{pr} \hat X_{n,m} =   \hat X_{n,m} +\hat X_{n+1,m} +\hat X_{n+2,m} +\hat X_{n,m+1} +\hat X_{n,m+2} + \hat X_{n+1,m+1} \label{e3-5b}
\eea
 of the vector field
 \bea
\hat X_{n,m} = \xi_{n,m} \partial_{x_{n,m}} + \tau_{n,m} \partial_{y_{n,m}} + \phi_{n,m} \partial_{u_{n,m}} \label{e3-4b}
\eea
 to the new set of independent
variables (\ref{f_1}, \ref{f_2}). We get
\bea 
\fl \mbox{pr} \hat X_{n,m} = \hat X_{n,m} + \sum_{(i,j)=0,1}\Big [\eta^{(x)}_{n+i,m+j} \partial_{h^x_{n+i,m+j}}  + \chi^{(x)}_{n+i,m+j} \partial_{\sigma^x_{n+i,m+j}}  + \eta^{(y)}_{n+i,m+j} \partial_{h^y_{n+i,m+j}}+\nonumber\\ 
\fl \qquad  + \chi^{(y)}_{n+i,m+j} \partial_{\sigma^y_{n+i,m+j}}\Big ]  +  \phi_{n,m}^{(1,x)} \partial_{D_x u_{n,m}}+\phi_{n,m}^{(1,y)}  \partial_{D_y u_{n,m}} +\phi_{n,m}^{(2,xx)} \partial_{[D_x]^2 u_{n,m}} + \nonumber\\ 
\fl \qquad  +\phi_{n,m}^{(2,xy)} \partial_{D_y D_x u_{n,m}}+ \phi_{n,m}^{(2,yy)} \partial_{[D_y]^2 u_{n,m}},\label{f_3}
\eea
where 
\bea \label{f_4}
\fl \eta^{(x)}_{n+i,m+j}=\xi_{n+1+i,m+j}-\xi_{n+i,m+j}, \;   \eta^{(y)}_{n+i,m+j}=\tau_{n+i,m+1+j}-\tau_{n+i,m+j},\\ \nonumber
\fl \chi^{(x)}_{n+i,m+j}=\xi_{n+i,m+1+j}-\xi_{n+i,m+j}, \,  \chi^{(y)}_{n+i,m+j}=\tau_{n+1+i,m+j}-\tau_{n+i,m+j}, \\ \nonumber
\fl \phi_{n,m}^{(1,x)} = D_x \phi_{n,m} - D_x u_{n,m} D_x \xi_{n,m}  - D_y u_{n,m} D_x \tau_{n,m} \\ \nonumber \fl \phi_{n,m}^{(1,y)} = D_y \phi_{n,m} - D_x u_{n,m} D_y \xi_{n,m}  - D_y u_{n,m} D_y \tau_{n,m},\\ \nonumber \fl \phi_{n,m}^{(2,xx)} = D_x \phi_{n,m}^{(1,x)}  - [D_x]^2 u_{n,m} D_x \xi_{n,m}  - D_y D_x u_{n,m} D_x \tau_{n,m}, \\ \nonumber 
\fl \phi_{n,m}^{(2,xy)} = D_x \phi_{n,m}^{(1,y)}  - D_x D_y u_{n,m} D_x \xi_{n,m}  - [D_y]^2 u_{n,m} D_x \tau_{n,m}, \\ \nonumber
\fl \phi_{n,m}^{(2,yy)} = D_y \phi_{n,m}^{(1,y)}  - D_x D_y u_{n,m} D_y \xi_{n,m}  - [D_y]^2 u_{n,m} D_y \tau_{n,m}.
\eea
In the continuous limit, when  $h^x_{n+i,m+j}$, $h^y_{n+i,m+j}$, $\sigma^x_{n+i,m+j}$ and $\sigma^y_{n+i,m+j}$   go to 0, $\eta^{(x)}_{n+i,m+j}$, $\eta^{(y)}_{n+i,m+j}$, $\chi^{(x)}_{n+i,m+j}$ and $\chi^{(y)}_{n+i,m+j}$ go also to 0 while $\phi_{n,m}^{(1,x)}$, $\phi_{n,m}^{(1,y)} $, $\phi_{n,m}^{(2,xx)} $, $\phi_{n,m}^{(2,xy)} $ and $\phi_{n,m}^{(2,yy)} $ go to the corresponding continuous prolongations.

Applying the infinitesimal generator (\ref{f_3}) onto (\ref{f_5}) we get that both  functions $\xi_{n,m}(x_{n,m},y_{n,m},u_{n,m})$ and $\tau_{n,m}(x_{n,m},y_{n,m},u_{n,m})$ must satisfy the discrete wave equations
\bea \label{f_6}
&&\xi_{n,m+1} - \xi_{n,m}-\xi_{n+1,m+1}+\xi_{n+1,m}=0, \\ \nonumber
&&\tau_{n,m+1} - \tau_{n,m}-\tau_{n+1,m+1}+\tau_{n+1,m}=0,
\eea
This is a constraint for  the symmetry coefficients if the Clairaut--Schwarz--Young theorem is to be satisfied, i.e.~(\ref{f_6})  are to be added to the determining equations if we want to have a lattice satisfying the Clairaut--Schwarz--Young theorem.  
In this case, if, for example, the difference equation for $u_{n,m}$ involves second order shifts like $u_{n+2,m}$ or $u_{n,m+2}$ so that $u_{n,m}$, $u_{n,m+1}$ and $u_{n+1,m}$ are independent variables, then from (\ref{f_6}) we get $\xi_{n,m}(x_{n,m},y_{n,m},u_{n,m})=\xi_{n,m}(x_{n,m}, y_{n,m})$. If the lattice equations in our scheme involve the points $x_{n+2,m}, y_{n+2,m}$ or $x_{n,m+2}, y_{n,m+2}$,  so that $x_{n,m}$, $x_{n,m+1}$, $x_{n+1,m}$, $y_{n,m}$, $y_{n,m+1}$ and $y_{n+1,m}$ are independent variables, then $\xi_{n,m}(x_{n,m},y_{n,m},u_{n,m})=\xi_{n,m}$. Then, as it is the case of the continuous wave equation,  the general  solution of (\ref{f_6}) is given by $\xi_{n,m}=f_n^{(x)}+g_m^{(x)}$ and $\tau_{n,m}=f_n^{(y)}+g_m^{(y)}$, i.e. the sum of an arbitrary function of $n$ and one of $m$.

It is worthwhile, in view of the application  to be carried out in next Section, to compute the lowest order discrete derivatives of monomials in $x$ and $y$:
 \bea \label{f_7}
 &&D_x x_{n,m}=1, \quad D_x y_{n,m}=0, \quad D_y x_{n,m} =0, \quad D_y y_{n,m}=1, \\ \nonumber
 && D_x x_{n,m}^2 = 2 x_{n,m} + \frac{h^y_{n,m} (h^x_{n,m})^2- \sigma^y_{n,m} (\sigma^x_{n,m})^2}{h^y_{n,m} h^x_{n,m} -\sigma^y_{n,m} \sigma^x_{n,m} } = 2 x_{n,m}+\Delta_{xx}^x ,\\ \nonumber
 && D_x x_{n,m} y_{n,m} =  y_{n,m} + \frac{h^y_{n,m} \sigma^y_{n,m} (h^x_{n,m} -  \sigma^x_{n,m} )}{h^y_{n,m} h^x_{n,m} -\sigma^y_{n,m} \sigma^x_{n,m} } =  y_{n,m} + \Delta_{xy}^x,\\ \nonumber
  && D_x  y_{n,m}^2 =  - \frac{h^y_{n,m} \sigma^y_{n,m} (h^y_{n,m} -  \sigma^y_{n,m} )}{h^y_{n,m} h^x_{n,m} -\sigma^y_{n,m} \sigma^x_{n,m} }=\Delta_{yy}^x,\\ \nonumber
&& D_y x_{n,m}^2 = -\frac{h^x_{n,m} \sigma^x_{n,m} (h^x_{n,m} -  \sigma^x_{n,m} )}{h^y_{n,m} h^x_{n,m} -\sigma^y_{n,m} \sigma^x_{n,m} }= \Delta_{xx}^y,\\ \nonumber
&& D_y x_{n,m} y_{n,m} = x_{n,m} + \frac{h^x_{n,m} \sigma^x_{n,m} (h^y_{n,m} -  \sigma^y_{n,m} )}{h^y_{n,m} h^x_{n,m} -\sigma^y_{n,m} \sigma^x_{n,m} }= x_{n,m} +\Delta_{xy}^y,\\ \nonumber
&& D_y y_{n,m}^2=2 y_{n,m} + \frac{h^x_{n,m} (h^y_{n,m} )^2- \sigma^x_{n,m} (\sigma^y_{n,m} )^2}{h^y_{n,m} h^x_{n,m} -\sigma^y_{n,m} \sigma^x_{n,m} }=2 y_{n,m} +\Delta_{yy}^y.
 \eea
 where the quantities $\Delta_{xx}^x$, $ \Delta_{xy}^x$, $\Delta_{yy}^x$, $\Delta_{xx}^y$, $\Delta_{xy}^y$ and $\Delta_{yy}^y$ 
  go to zero in the continuous limit when $h$ and $\sigma$ go to zero.

\section{Example: the potential Burgers equation}
The Burgers equation
\bea \label{b1}
u_t=\nu u_{xx}+u u_x,
\eea
 a very well known partial differential equation, appears as a simplification of the Navier--Stokes equation and has been studied from many, if not all, points of view \cite{burgers}. It was proposed as a model for a viscous fluid, with a viscosity parameter $\nu$.  When the viscosity parameter $\nu$ is set equal to zero, the Burgers equation degenerates into a quasilinear first order equation which is the prototype of a class of equations which presents nonlinear phenomena such as shock waves. In fact, the limit $\nu\to 0$ allows the study of these shock wave solutions as limits of the solutions of the viscous Burgers equation. In particular, although the inviscid Burgers equation has an infinite dimensional group of symmetries (being a first order equation), the limit of the symmetry group of the viscous Burgers equations provides a subgroup of the whole group of symmetries of the inviscid Burgers equation which is a  useful tool in the study of the equation and in particular  its discretization using invariant techniques.
 
Several invariant discretization approaches have been proposed  to construct explicit numerical schemes for finding numerical solutions on invariant lattices \cite{BN13,Ki08}. In some of these works, explicit comparison has been made, showing the higher accuracy and stability of these methods \cite{CH10}.

It is not our intention in this paper  to present this kind of numerical results but rather to prove the possibility of constructing in an easy way such invariant  discrete schemes and  discuss the properties of a lattice satisfying the Clairaut--Schwarz--Young theorem from the numerical point of view. For the control of the numerical calculations we will study the time evolution of the initial condition provided by exact solutions of the Burgers equation. To simplify the presentation and with no loss of generality we will go over to consider the {\bf potential Burgers} equation as this is point transformable into the linear heat equation for which many exact solutions are known.

Let us construct using the formulas introduced in the previous section the discrete scheme which preserves the point symmetries of the potential Burgers equation
\bea \label{k_0}
u_y - u_{xx} -  u_x^2 =0.
\eea 
The point symmetries of (\ref{k_0}) are \cite{olver}
\bea \label{k_1}
\hat V_1 &=& \partial_x, \quad \hat V_2 = \partial_y, \quad \hat V_3 = \partial_u, \quad \hat V_4 = x \partial_x + 2 y \partial_y, \\ \nonumber
\hat V_5 &=& 2 y \partial_x - x \partial_u, \quad \hat V_6=4 y x \partial_x + 4 y^2 \partial_y -(x^2 + 2 y)\partial_u, \\ \nonumber 
 \hat V_{\alpha} &=& \alpha(x,y) e^{-u} \partial_u,
\eea
where the function $\alpha(x,y)$ satisfies the heat equation 
\bea \label{21b}
\alpha_y=\alpha_{xx}.
\eea
The infinitesimal generator $\hat V_{\alpha}$ is the one responsible for the linearizability of the potential Burgers equation as it provides its linearizing transformation
\bea \label{21c}
\alpha=e^{u}.
\eea 
A function $F(x,y,u,u_x,u_y, u_{xx})$ is invariant under the infinitesimal generators $\hat V_1$, $\hat V_2$, $\hat V_3$  $\hat V_4$, $\hat V_5$   if it depends on 
\bea \label{k_2}
I^{(1)}=\frac{u_y - u_{x}^2}{u_{xx}}.
\eea
It is then easy to see that (\ref{k_2}) is weakly invariant also under  $\hat V_6$ and $ \hat V_{\alpha}$, i.e.
\bea \nonumber
\mbox{pr}^{(2)}\hat V_6 \, I^{(1)} &=& \frac{2}{u_{xx}} ( I^{(1)}-1),\\ \nonumber
\mbox{pr}^{(2)}\hat V_{\alpha} \, I^{(1)} &=& -\frac{e^{-u}}{u_{xx}}\left (\alpha u_x^2 +2 \alpha_x u_x +\alpha_y \right )  ( I^{(1)}-1).
\eea
The potential Burgers equation (\ref{k_0}) is then given by
\bea \label{k_3}
I^{(1)}=1
\eea
Taking into account the discrete prolongation (\ref{f_3}) and the definition of the infinitesimal coefficients (\ref{f_4}) we can construct the discrete prolongation of the vector fields (\ref{k_1}). 
It is easy to show that the conditions (\ref{f_6}) are satisfied for all symmetries (\ref{k_1}) except for $\hat V_6$ for which 
$$
\xi_{n,m+1} - \xi_{n,m}-\xi_{n+1,m+1}+\xi_{n+1,m}=-4[h^x_{n,m} h^y_{n,m}+\sigma^x_{n,m} \sigma^y_{n,m}] \ne 0$$
and
$$\tau_{n,m+1} - \tau_{n,m}-\tau_{n+1,m+1}+\tau_{n+1,m}=-8h^x_{n,m} \sigma^x_{n,m}\ne 0.
$$
Since we are presently interested in analyzing the role of the Schwarz condition on the construction of lattices and its consequences in numerical computations, we will not consider the symmetry with generator $\hat{V}_6$ in the following, that is, we restrict ourselves to a sugbroup of the whole symmetry group (note that we have also disregard in our analysis the operator $V_{\alpha}$, which, as we have remarked above, is related to the linearization of the equation under study).

Taking into account (\ref{f_7}) we have:
\bea \label{k_4}
\fl \mbox{pr}^d \hat V_1 = \partial_{x_{n,m}}, \quad \mbox{pr}^d \hat V_2 = \partial_{y_{n,m}}, \quad \mbox{pr}^d \hat V_3 = \partial_{u_{n,m}}, 
\\ \nonumber
\fl \mbox{pr}^d \hat V_4 = x_{n,m} \partial_{x_{n,m}} +2 y_{n,m} \partial_{y_{n,m}} + h^x_{n,m} \partial_{h^x_{n,m}} + h^x_{n+1,m} \partial_{h^x_{n+1,m}} +h^x_{n,m+1} \partial_{h^x_{n,m+1}}+ \\ \nonumber 
\fl \qquad +\sigma^x_{n,m} \partial_{\sigma^x_{n,m}}+\sigma^x_{n,m+1} \partial_{\sigma^x_{n,m+1}} +  2 \left ( h^y_{n,m} \partial_{h^y_{n,m}} + h^y_{n+1,m} \partial_{h^y_{n+1,m}} +\right .\\ \nonumber 
 \fl \qquad +h^y_{n,m+1} \partial_{h^y_{n,m+1}}+ \sigma^y_{n,m} \partial_{\sigma^y_{n,m}}+\sigma^y_{n+1,m} \partial_{\sigma^y_{n+1,m}}\left. \right )- D_x u_{n,m} \partial_{D_x u_{n,m}} -\\ \nonumber 
\fl \qquad -2 D_y u_{n,m} \partial_{D_y u_{n,m}} - 2 [D_x]^2 u_{n,m} \partial_{D_x^2 u_{n,m}} ,
\\ \nonumber
\fl \mbox{pr}^d \hat V_5 = 2 y_{n,m} \partial_{x_{n,m}} -x_{n,m} \partial{u_{n,m}} +2 \sigma^y_{n,m} \partial_{h^x_{n,m}} + 2 \sigma^y_{n+1,m} \partial_{h^x_{n+1,m}} + \\ \nonumber 
\fl \qquad +2 \left ( h^y_{n+1,m} + \sigma^y_{n,m} - h^y_{n,m} \right )  \partial_{h^x_{n,m+1}}+ 2 h^y_{n,m} \partial_{\sigma^x_{n,m}}+2 h^y_{n,m+1} \partial_{\sigma^x_{n,m+1}} -  \\ \nonumber 
\fl \qquad - \partial_{D_x u_{n,m}} -2  D_x u_{n,m} \partial_{D_y u_{n,m}}.
\eea
The commutation table of this algebra appears in Table \ref{tab} (it is, obviously the same as in the continuous case).

\begin{table}[h]
\begin{center}
\begin{tabular}{l|lllll}
       & $V_1$ & $V_2$ &  $V_3$ & $V_4$ & $V_5$ \\
\hline
 $V_1$ & $0$ & $0$ &  $0$ & $V_1$ & $-V_3$ \\
 $V_2$ & $0$ & $0$ &  $0$ & $2V_2$ & $2V_1$ \\
 $V_3$ & $0$ & $0$ &  $0$ & $0$ & $0$ \\
 $V_4$ & $-V_1$ & $-2V_2$ &  $0$ & $0$ & $V_5$ \\
 $V_5$ & $V_3$ & $-2V_1$ &  $0$ & $-V_5$ & $0$
\end{tabular}
\caption{\label{tab} Commutation table of the discrete invariance algebra}
\end{center}
\end{table}

It is immediate to see that a discrete potential Burgers  equation preserving the Lie algebra of (\ref{k_0}) given by the generators $\hat V_1$, $\hat V_2$, $\hat V_3$  $\hat V_4$, $\hat V_5$ is: 
\bea \label{k_5}
\fl \mathcal{I}^{(1)}=\frac{[D_{x}]^2 u_{n,m}}{D_y u_{n,m} - (D_x u_{n,m})^2}=1, \; \mbox{ i.e.}\; D_y u_{n,m} - [D_x]^2 u_{n,m} - (D_x u_{n,m})^2 =0.
\eea
The continuous limit of (\ref{k_5}) is trivially given by (\ref{k_0}) when $h^x$,  $h^y$, $\sigma^x$ and $\sigma^y$  go to zero preserving the structure of the lattice. Eq. (\ref{k_5}) involves 6 lattice points centered around ($n,m$), i.e. ($n,m$), ($n+1,m$), ($n,m+1$), ($n+2,m$), ($n+1,m+1$), ($n,m+2$). It explicitly reads:
\bea \label{k_5a}
&&\frac{-\sigma_{m}^x \left(u_{n+1,m}-u_{n,m} \right) 
+h_{n}^x \left( u_{n,m+1}-u_{n,m}\right)}
{h_{n}^xh_{m}^y-\sigma_{m}^x\sigma_{n}^y}
\\ \nonumber &&
- \left[ h_{m}^y\left( 
\frac {h_{m}^y \left( u_{n+2,m}-u_{n+1,m} \right) -
\sigma_{n+1}^y \left( u_{n+1,m+1}-u_{n+1,m} \right) }
{h_{n+1}^xh_{m}^y-\sigma_{m}^x\sigma_{n+1}^y} 
\right.\right.
\\ \nonumber && 
\left. -\frac {h_{m}^y \left( u_{n+1,m}-u_{n,m} \right) 
-\sigma_{n}^y \left( u_{n,m+1}-u_{n,m} \right) }
{h_{n}^xh_{m}^y-\sigma_{m}^x\sigma_{n}^y} 
\right) 
\\ \nonumber &&
-\sigma_{n}^y\left( \frac {h_{m+1}^y \left( u_{n+1,m+1}-u_{n,m+1}\right) 
 -\sigma_{n}^y \left( u_{n,m+2}-u_{n,m+1} \right) }
{h_{n}^xh_{m+1}^y-\sigma_{m+1}^x\sigma_{n}^y} 
\right.
\\ \nonumber && 
\left.\left. -\frac {h_{m}^y \left( u_{n+1,m}-u_{n,m} \right) 
-\sigma_{n}^y \left( u_{n,m+1}-u_{n,m} \right) }
{h_{n}^xh_{m}^y-\sigma_{m}^x\sigma_{n}^y} 
\right)  \right]  
\left( h_{n}^xh_{m}^y-\sigma_{m}^x\sigma_{n}^y\right)^{-1}
\\ \nonumber && 
- \frac {\left( h_{m}^y \left( u_{n+1,m}-u_{n,m} \right) 
-\sigma_{n}^y \left( u_{n,m+1}-u_{n,m} \right) \right) ^{2}}
{ \left( h_{n}^xh_{m}^y-\sigma_{m}^x\sigma_{n}^y \right) ^{2}}=0
\eea

To complete the difference scheme we have to associate to it a lattice equation which preserve the symmetries (\ref{k_4}) or part of them.
The complete list of discrete invariant of  (\ref{k_4}), obtained as usual, as solutions of the equations $\mbox{pr}^d \hat V_i(\mathcal{K})=0$, are:
\bea \label{k_6}
\fl \mathcal{K}_1 = &\frac{h^y_{n+1,m}}{h^y_{n,m}},\quad
\mathcal{K}_2 = \frac{h^y_{n,m+1}}{h^y_{n,m}},\quad
\mathcal{K}_3=  \frac{\sigma^y_{n,m}}{h^y_{n,m}},\quad
\mathcal{K}_4=  \frac{\sigma^y_{n+1,m}}{h^y_{n,m}},\\ \nonumber
\fl\mathcal{K}_5=&\frac{1} { (h^y_{n,m})^{3/2}} \left(h^x_{n,m}  h^y_{n,m}- \sigma^x_{n,m}  \sigma^y_{n,m} \right),\quad \mathcal{K}_6=  \frac{1}{(h^y_{n,m})^{3/2}}\left(   h^y_{n,m+1}\sigma^x_{n,m}- h^y_{n,m }  \sigma^x_{n,m+1}\right)
,\\ \nonumber
\fl\mathcal{K}_7= & 
\frac{1}{ (h^y_{n,m})^{3/2}} \left(h^x_{n,m}(h^y_{n+1,m}- h^y_{n,m})- \sigma^y_{n,m}(h^x_{n,m+1} - h^x_{n,m})\right),\\  \nonumber
\fl\mathcal{K}_8=&   
\frac{1}{ (h^y_{n,m})^{3/2}}\left( h^x_{n,m}  \sigma^y_{n+1,m}- h^x_{n+1,m}  \sigma^y_{n,m}\right),\quad
\mathcal{K}_9=\frac{1}{(h^y_{n,m})^{1/2}}\left(h^x_{n,m}+2  \sigma^y_{n,m}    D_xu_{n,m} \right),\\  \nonumber
\fl\mathcal{K}_{10}=& \frac{ \left(  {D}_yu_{n,m}-  (D_xu_{n,m})^2\right)}{[D_x]^2 u_{n,m}}.
\eea

Our first choice for the difference scheme is an orthogonal cartesian lattice given, in the coordinates $h_{n,m}$ and $\sigma_{n,m}$ by 
\bea \label{1}
h^y_{n,m}=b, \qquad \sigma^y_{n,m}=0,\qquad h^x_{n,m}=a, \qquad \sigma^x_{n,m}=0,
\eea
where $a$ and $b$ are arbitrary constants, which in the continuous limit go to zero. This lattice, which is clearly invariant under all the symmetries which satisfy the commutativity constraint, corresponds to the Lie point symmetry infinitesimal generators
\bea \label{1a}
\xi_{n,m}=\tau_{n,m}=1
\eea
These generators comply with the constraints in (\ref{f_6}) and the orthogonal lattice (\ref{1}) satisfies the Clairaut--Schwarz--Young theorem.

To carry out a comparison between this invariant Schwarzian lattice with other lattices which lack these properties, we will consider an exponential non Schwarzian lattice as given in \cite{LRI} by
\bea \label{2}
y_{n,m}=b\, m + b_0, \qquad x_{n,m}=(1+c)^m (a\, n + a_0),
\eea
where $a$, $a_0$, $b$, $b_0$ and $c$ are arbitrary constants. $b$ and $a$ are the lattice spacing and $c$ is a dilation parameter which, when set equal to zero reduces this lattice to an orthogonal lattice. This lattice corresponds to the Lie point symmetry infinitesimal generators
\bea \label{2a}
\xi_{n,m}=(1+c)^m (k_1 n +k_2) + k_0 \, x_{n,m}, \qquad \tau_{n,m}=k_3
\eea
which clearly do not satisfy (\ref{f_6}). The non Schwarzian property of this lattice can also be seen by considering the lattice differences
\bea \label{2c}
h^y_{n,m}=b, \quad \sigma^y_{n,m}=0,\quad h^x_{n,m}=(1+c)^m a, \quad \sigma^x_{n,m}=c(1+c)^m a \, n.
\eea
and comparing them with (\ref{f_5}). This non Schwarzian lattice is not invariant under the potential Burgers symmetry group (\ref{k_1}); $\mathcal{K}_5$, $\mathcal{K}_6$, $\mathcal{K}_7$ and $\mathcal{K}_9$ are non constants.

In the following Section we will provide a numerical test about the possible importance of the Clairaut--Schwarz--Young theorem, by studying numerically the evolution provided by the symmetry preserving discretized potential Burgers equation (\ref{k_5a}) on the two different lattices we introduced above, i.e.  the Schwarzian and non Schwarzian lattices (\ref{1}, \ref{2c}).

\section{Numerical calculation results for the discrete potential Burgers equation}

In order to compare the precision and accuracy of different lattices used to compute numerically the solution of the potential Burgers equation, we will construct two of its exact solutions (associated to different symmetries of the heat equation) and use them as initial conditions for the evolution of the Burgers map (27). In fact, we will not do a direct comparison of the evolution of this map in the two lattices (29) and (33) but compare the solution of the map in each lattice with the exact solution of the continuous equation.

We introduce two invariant solutions of (\ref{21b}) and transform them into solutions of the potential Burgers equation \cite{LRI} using (\ref{21c}). Starting from the traveling wave solution of the heat equation (invariant under a combination of $x$ and $y$ translations) we get as a simple solution, bounded for $x \in \mathbf{R}^+$ for each $y$,
\bea \label{3}
f_1(x,y)=\log (1+e^{-(x-y)}).
\eea
A second exact solution (the fundamental solution)  is given by the Galilei--invariant solution of the heat equation 
\bea \label{4}
f_2(x,y) =\log \left(1+\frac{e^{-x^2/(4y)}}{\sqrt{y}}\right).
\eea
To compare the results given by the evolution of the map on a given lattice to the evolution given by the exact solution we introduce a global estimator, the usual (relative) distance in the discrete analog of the $L^2$ space:
\begin{equation}
\chi_{\mathrm{lattice}}(f)=\sqrt{\frac{\sum_{n,m}(f_{n,m}^{\mathrm{lattice}}-f_{n,m})^2}{\sum_{n,m}f_{n,m}^2}}
\end{equation} 
where $f_{n,m}$  and $f^{\mathrm{lattice}}_{n,m}$ are  the values of the exact solution  and the numerical one, respectively, computed in the points of the   lattices (\ref{1}, \ref{2c}).

\begin{figure}[h]
\begin{center}$
\begin{array}{cc}
(1) \includegraphics[width=1.8in]{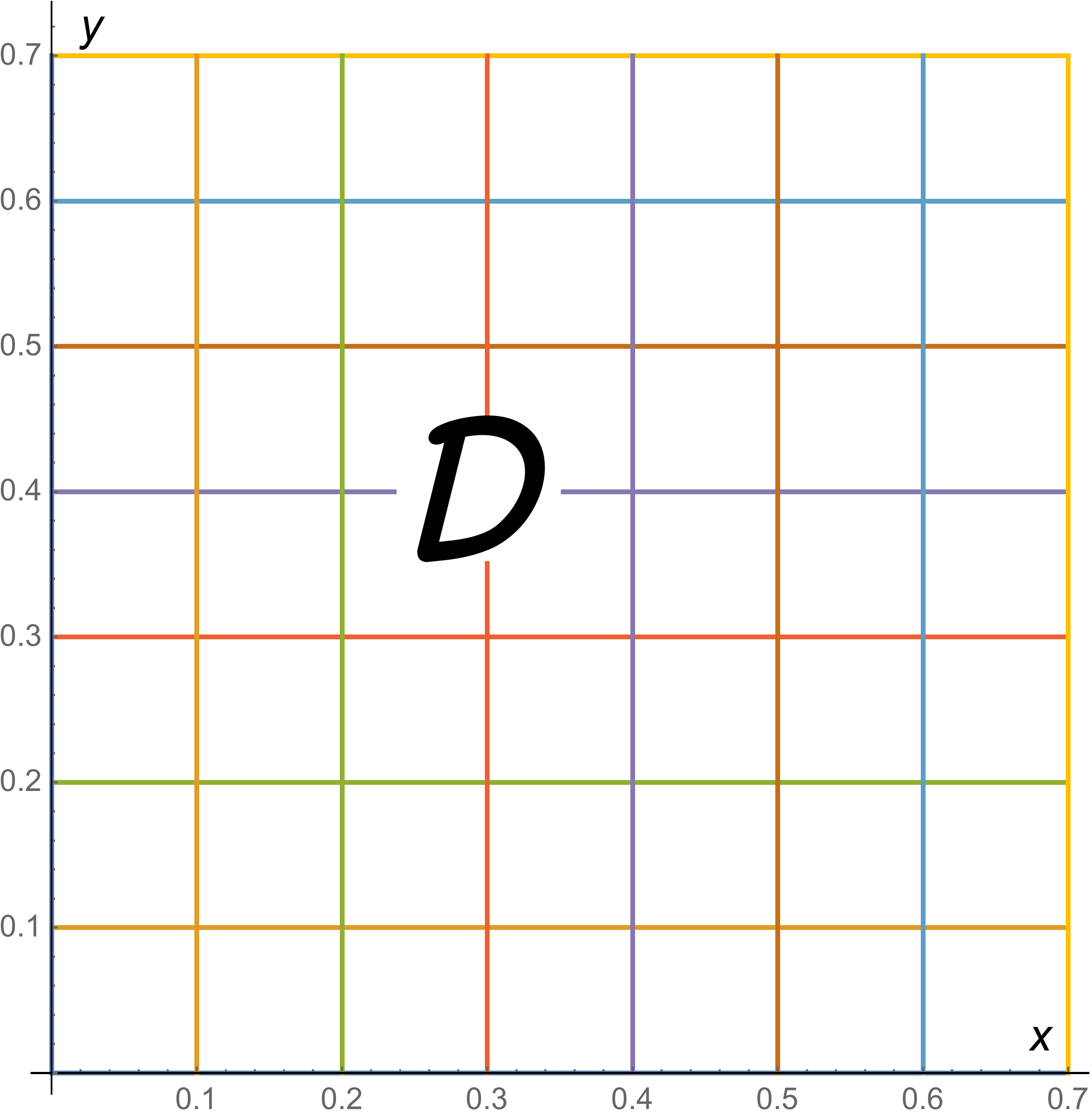} &
(2) \includegraphics[width=2.8in]{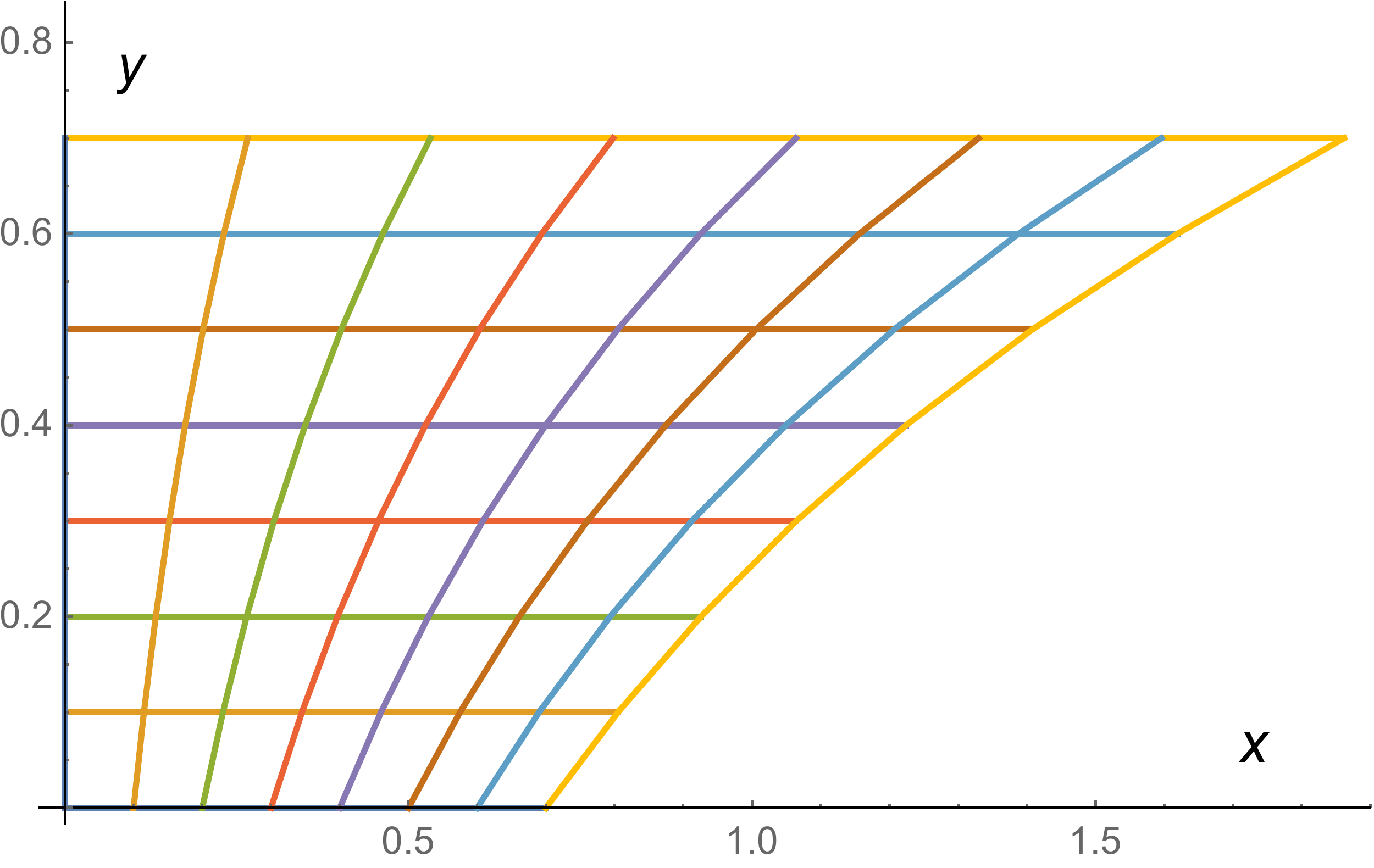} \\[5pt] 
(3) \includegraphics[width=2.0in]{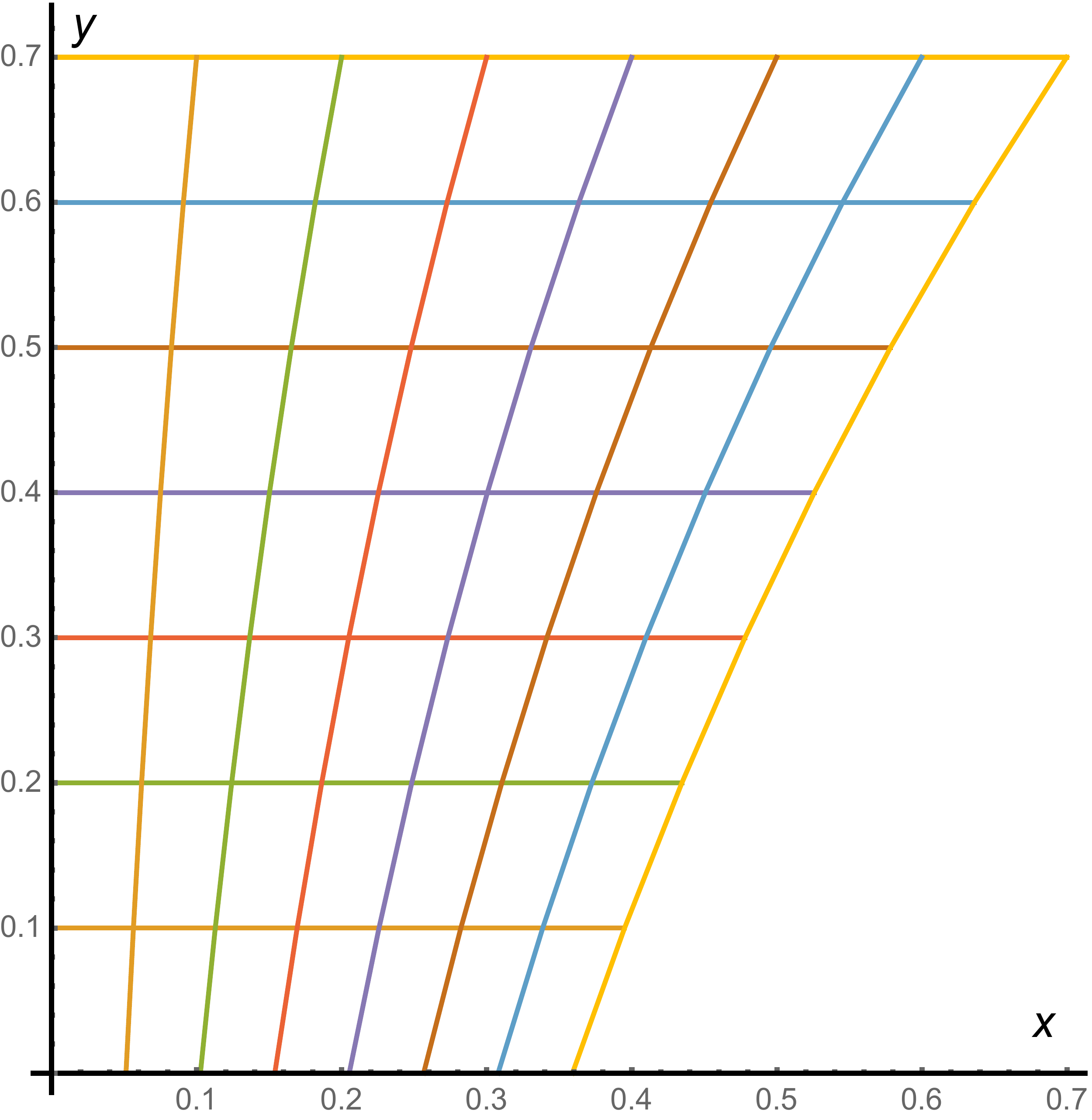} & 
(4) \includegraphics[width=2.0in]{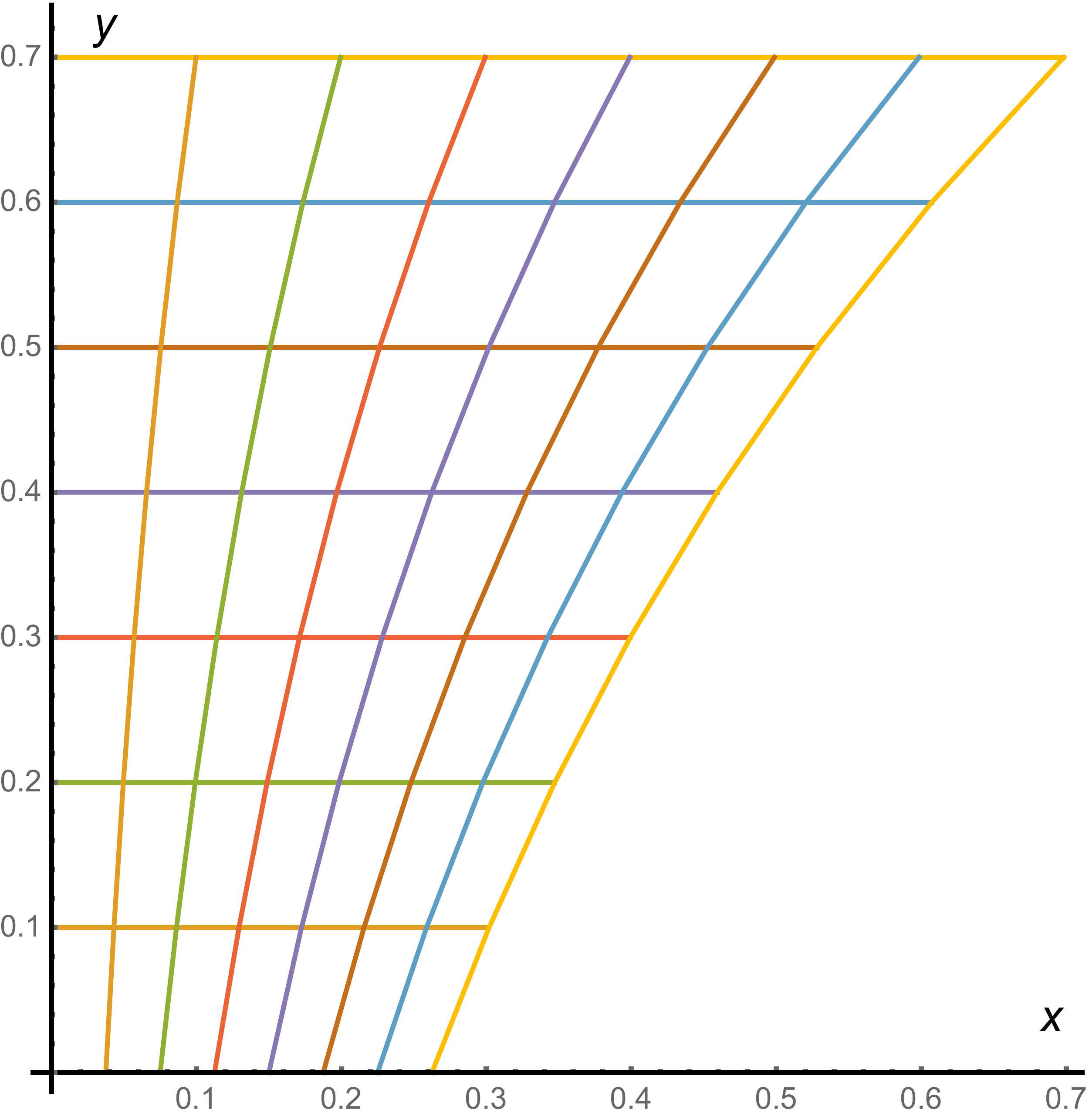}  
\end{array}$
\caption{\label{fig1} $(1)$ Orthogonal lattice: $a=0.1$, $b=0.1$. $(2)$ Exponential lattice: $a=0.1$, $b=0.1$, $c=0.15$. $(3)$ Exponential lattice, $a=0.0513$, $b=0.1$, $c=0.1$. $(4)$ Exponential lattice, $a=0.0375$, $b=0.1$, $c=0.15$. Dimension: $8\times 8$ points.}
\end{center}
\end{figure}

Since our intention is to compare the lattices, we will not insist on improving the precision and accuracy of the solution by  modifying in an optimal way the parameters involved in the computation. For the orthogonal lattice we will take (for $f_1$ and $f_2$) $a=b=0.1$ in a square $\mathcal{D}$ of $8\times 8$ points in the $x,y$ plane. We will use the same number of points in all the cases we will consider, in order to keep as far as possible the same round off errors due to machine precision. Augmenting the number of points enlarge numerical instabilities which could be reduced by increasing the precision of the calculations at the cost of the time of calculus. These instabilities are {\it non--physical} and we decide to avoid them by reducing appropriately the number of points. The region  covered by the lattice in the two cases (orthogonal and exponential) can be different for the same lattice spacing. See for example in fig.~\ref{fig1} the cases (1) and (2). In the exponential lattice we will consider two different situations: (i) the spacing is the same as in the orthogonal case ($a=b=0.1$), and thus for the exponential lattice the  region is deformed and enlarged with respect to the square $\mathcal{D}$ considered in the orthogonal lattice, (ii) we will modify $a$ in such a way that the 64 points of the lattice are inside the  square $\mathcal{D}$. Let us notice that in this case part of $\mathcal{D}$ is not covered by the lattice, and this may create problems as we will see later.

The exponential lattice has a parameter $c$ controlling the dilation of the $x$ variable. As we said above, when $c=0$ the lattice is orthogonal. We have considered in these numerical calculations two cases, $c=0.1$ and $c=0.15$, to compare the different behavior of the exponential lattice when it turns into an orthogonal one. In these two cases, the parameter $a$ is taken as $a=0.0375$ and $a=0.0513$ respectively, when we keep the 64 point of the lattice inside $\mathcal{D}$ (see fig.~\ref{fig1} for a graphical description of the four lattices under study).

The $\chi$ estimator is given in Table \ref{tab2} and gives raise to the following conclusions:
\begin{enumerate}
\item The orthogonal lattice (Schwarzian lattice) provides better results than the exponential one (non Schwarzian lattice) in all cases except one. 
\item The results for the exponential lattice with different values of $c$ show that the approximation is better when the lattice is closer to a Schwarzian lattice (recall that when $c\to 0$ the exponential lattice becomes orthogonal), for the two solutions considered.
\item The value of the $\chi$ estimator for the traveling wave solution in the exponential lattice when $a=0.0513$ and $c=0.1$  is lower than the corresponding value for the orthogonal case. In this case (case 3 of Fig.1) the lattice is close to the orthogonal one but there is a region in $\mathcal{D}$ which is not covered by the exponential lattice. This is the region where the round off errors   the computer makes in calculating the points  in the orthogonal lattice are greater.  This is the reason for this unsatisfactory value. When $c=0.15$ the lattice is far from the orthogonal one and thus $\chi$ is greater than in the orthogonal case.
\end{enumerate}

\begin{table}[h]
\begin{center}
\begin{tabular}{c|c|c|c}
& $\chi_{\mathrm{ort}}$ &  $\chi_{\mathrm{exp}}$, {\small $c=0.1$}  & $\chi_{\mathrm{exp}}$, {\small $c=0.15$} \\
\hline
$f_1$ &  $(1)$ $0.01267$ & $(5)$ $0.01437$ & $(2)$ $0.01651$\\
 &   & $(3)$ $0.01147$ & $(4)$ $0.01408$ \\
\hline
$f_2$ & $(1)$ $0.00249$ &  $(5)$ $0.00430$ & $(2)$ $0.00610$\\
 &   & $(3)$ $0.00642$  & $(4)$ $0.00913$ \\
\hline
\end{tabular}
\caption{\label{tab2}$\chi$ estimator values for the functions $f_1$ and $f_2$ given by (\ref{3}) and (\ref{4}), respectively. $(1)$: $a=0.1$, lattice $(1)$ in fig.~\ref{fig1}. $(2)$: $a=0.1$, lattice $(2)$ in fig.~\ref{fig1}. $(3)$: $a=0.0513$, lattice $(3)$ in fig.~\ref{fig1}. $(4)$: $a=0.0375$, lattice $(4)$ in fig.~\ref{fig1}. $(5)$:  $a=0.1$, the exponential lattice is similar to the lattice $(2)$ in fig.~\ref{fig1}. }
\end{center}
\end{table}

\section{Conclusions}
In this work we have shown that also in the case of partial difference equation we can introduce a set of variables which are in one to one correspondence with the grid points when we substitute them by the lattice differences and the derivatives on the lattice of the dependent function. This correspondence allows us to write down the invariance equations by using  only the  knowledge of the continuous invariants. 

The numerical calculation we carried out in Section 4 shows that the Schwarzian property seems important in providing better numerical results. 

More examples should be done, both to understand the stability of the models we construct in this way and the analysis of the lattices in the Schwarzian and non Schwarzian case. There might be boundary value problems where a non Schwarzian lattice adapted to the geometry of the problem could be better than a Schwarzian one.

\ack 
DL has been partly supported by the Italian Ministry of Education and Research, 2010 PRIN {\it Continuous and discrete nonlinear integrable evolutions: from water waves to symplectic maps} and   by INFN   IS-CSN4 {\it Mathematical Methods of Nonlinear Physics}. DL thanks the Departamento de F\'{\i}sica Te\'orica II of the Complutense University in Madrid for its hospitality.
 MAR was supported by the Spanish MINECO under project  FIS2015-63966. The authors would like to thank the referees for their useful comments.

\end{document}